\documentstyle[twoside,fleqn,espcrc2]{article}

\def \lta {\mathrel{\vcenter{\hbox{$<$}\nointerlineskip\hbox{$\sim$}}}}

\newcommand{\AmS}{{\protect\the\textfont2
  A\kern-.1667em\lower.5ex\hbox{M}\kern-.125emS}}

\title{Quintessence, Supersymmetry and Inflation}

\author{Francesca Rosati\address{SISSA/ISAS, via Beirut 2-4, I-34013
    Trieste, ITALY \\ and INFN, Sezione di Trieste,  
    Padriciano 99, I-34012 Trieste, ITALY}\thanks{Work partially 
    supported under TMR contract ``BSM'' n$^o$ FMRX-CT960090}}
       
\begin{document}

\begin{abstract}
Recent data point in the direction of a $\Lambda$ dominated universe.
We briefly review ``Quintessence'' as a model for a dynamical 
cosmological term and analyse the role of Susy QCD as a
possible particle physics candidate.
The multiscalar content of the theory is fully taken into account
and interaction with other cosmological fields is discussed.
Finally, the possibility of constructing  a unified scheme for
quintessence and inflation is mentioned.
\end{abstract}

\maketitle

\section{INTRODUCTION}

The very last years have witnessed growing interest in cosmological models 
with $\Omega_m \sim 1/3$ and $\Omega_\Lambda \sim 2/3$, following
the most recent observational data (see for example  \cite{data}
and references therein). 
A very promising candidate for a dynamical cosmological constant 
$\Lambda(t)$ is a ``quintessential'' scalar field presently rolling down
its potential \cite{quint}, for which particle physics models
have also been proposed \cite{bin,noi,mod2}. 
Several ways of constraining these models from observations are 
under investigation \cite{det}. 

The study of scalar field cosmologies has 
shown \cite{scalcosmo} that for certain potentials
there exist attractor solutions that can be of the ``scaling''
\cite{scale} or ``tracker'' \cite{track} type; that means that
for a wide range of initial conditions the scalar field will rapidly
join a well defined late time behavior. 

On the other hand, the investigation of quintessence models from 
the particle physics point of view presents two classes of problems: 
the construction of a field theory with the required scalar 
potential and the interaction of the quintessence field with the 
standard model (SM) fields \cite{int}.  
The former problem was first considered by Bin\'{e}truy \cite{bin}, 
who pointed out that scalar inverse power law potentials (required by
the ``tracking'' condition) appear in supersymmetric QCD theories
with $N_{c}$ colors and $N_{f}<N_{c}$ flavors \cite{SQCD}. 
The second seems the toughest. Indeed the quintessence field today has 
typically a mass of order $H_{0}\sim 10^{-33}$eV. 
Then, in general, it would mediate long range interactions of gravitational 
strength, which are phenomenologically unacceptable. 

Going on with the analysis, another very interesting question is
whether it is possible to construct a succesful common scheme for the
two cosmological mechanisms involving rolling scalar fields, $i.e.$
quintessence and inflation.
This perspective has the appealing feature of providing a
unified view  of the past and recent history of the universe, 
but can also ``cure'' some weak points that the two mechanisms taken
separately have. Indeed, inflation could provide the initial
conditions for quintessence without any need to fix them by hand,
and quintessence could hope to give some more hints in constraining the
inflaton potential on observational grounds.

\section{QUINTESSENCE}

\subsection{The cosmological attractors}

Consider a cosmological scalar field $Q$, with potential $V(Q)$
evolving according to
\begin{equation}
\ddot{Q}+3H\dot{Q}+\frac{\partial V}{\partial Q}\ =\ 0  
\end{equation}
and whose equation of state is given by
\begin{equation}
w_{Q}=\frac{ \dot{Q}^{2}/2 -V(Q)}{ \dot{Q}^{2}/2 +V(Q)} \;\; ,
\end{equation}
with $H^{2}=8\pi /3M_P^{2}\ (\rho _{m}+\rho _{r}+\rho _{Q})$, where
$M_P$ is the Planck mass, $\rho _{m(r)}$ is the matter (radiation)
energy density, and $\rho_{Q}$ is the quintessence field energy. 
If $\rho _{Q}\ll \rho _{B}$, where $\rho _{B}$ is the energy
density of the dominant background (radiation or matter), the
attractor can be studied analytically. 

In the case of an exponential potential, $V \sim \exp {(-Q)}$ the
solution $Q\sim \ln {t}$ is, under very general conditions, a ``scaling''
attractor in phase space characterized by $\rho _{Q}/\rho _{B}\sim
{\rm const}$ \cite{scale}. 
This could potentially solve the so called ``cosmic coincidence'' problem, 
providing a dynamical explanation for the order of magnitude equality 
between matter and scalar field energy today.  
Unfortunately, the equation of state for this attractor is $w_{Q}=w_{B}$, 
which cannot explain the acceleration of the universe neither during RD 
($ w_{rad}=1/3$) nor during MD ($w_m=0$).  
Moreover, Big Bang nucleosynthesis constrain the field
energy density to values much smaller than the required $ \sim 2/3$
\cite{scalcosmo,scale}. 

If instead an inverse power-law potential is considered,
$V=M^{4+\alpha }Q^{-\alpha }$, with $\alpha >0$, the attractor
solution is $Q\sim t^{1-n/m}$, where $n=3(w_{Q}+1)$ and $m=3(w_{B}+1)$.
The equation of state turns out to be 
$w_{Q}= (\alpha w_B -2)/(\alpha+2)$,
which is always negative during MD. 
The ratio of the energies is no longer constant but scales as\ 
$\rho _{Q}/\rho_{B}\sim a^{m-n}$ thus growing during the cosmological 
evolution, since $n$ $<m$.  
$\rho _{Q}$ could then have been safely small during
nucleosynthesis and have grown lately up to the phenomenologically
interesting values.
These solutions are good candidates for
quintessence and have been denominated ``trackers'' in the literature
\cite{scalcosmo,track}. 
There are two main qualitative ways through which the attractor can be
joined (see \cite{track} for details). 
If the initial conditions for the scalar field $Q$ are such that 
$\rho^{0}_{cr} \leq \rho^{in}_{Q} \leq \rho^{in}_{\sc tr}$ 
({\it undershoot} case), it will remain ``freezed'' until 
$\rho_{Q} \sim \rho_{\sc tr}$ and then start to scale 
as the tracker. 
If, instead, initially $\rho^{in}_{\sc tr} \leq 
\rho^{in}_{Q} \leq \rho^{in}_{\sc b}$ 
({\it overshoot} case) then $Q$ will pass through 
a phase of kinetic energy domination before remaining freezed and 
eventually join the attractor.

The inverse power-law potential does not improve the cosmic
coincidence problem with respect to the cosmological constant
case. Indeed, the scale $M $ has to be fixed from the requirement that
the scalar energy density today is exactly what is needed. This
corresponds to choosing the desired tracker path.  An important
difference exists in this case though.  
The initial conditions for the physical variable $\rho _{Q}$ can vary 
over many tens of orders of magnitude (between the present critical
energy density and the initial background energy density), depending on the 
initial time, and will anyway end on the tracker path before the present 
epoch, due to the presence of an attractor in phase space \cite{track}.  
On the contrary, in the cosmological constant case, the physical variable
$\rho _{\Lambda }$ is fixed once for all at the beginning. This allows
us to say that in the quintessence case the fine-tuning issue, even if
still far from solved, is at least weakened. 

\subsection{The tracker solution in Susy QCD}

As already noted by Bin\`{e}truy \cite{bin}, supersymmetric QCD
theories with $N_{c}$ colors and $N_{f}<N_{c}$ flavors \cite{SQCD} may
give an explicit realization of a model for quintessence with an
inverse power law scalar potential. 

The matter content of the theory is given by the chiral superfields
$Q_{i}$ and $\overline{Q}_{i}$ ($i=1\ldots N_{f}$) transforming
according to the $ N_{c}$ and $\overline{N}_{c}$ representations of
$SU(N_c)$, respectively.  In the following, the same symbols will be
used for the superfields $Q_{i}$, $\overline{Q}_{i}$, and their scalar
components. 

Supersymmetry and anomaly-free global symmetries constrain the
superpotential to the unique {\it exact} form
\begin{equation}
W=(N_{c}-N_{f})\left( \frac{\Lambda ^{(3N_{c}-N_{f})}}{{\rm
det}T}\right) ^{ \frac{1}{N_{c}-N_{f}}} \label{superpot}
\end{equation}
where the gauge-invariant matrix superfield $T_{ij}=Q_{i}\cdot
\overline{Q}_{j}$ appears. $\Lambda $ is the only mass scale of the
theory.  
It is the supersymmetric analogue of $\Lambda _{QCD}$, the
renormalization group invariant scale at which the gauge coupling of
$SU(N_{c})$ becomes non-perturbative. As long as scalar field values
$Q_{i},\overline{Q}_{i}\gg $ $\Lambda $ are considered, the theory is
in the weak coupling regime and the canonical form for the K\"{a}hler
potential may be assumed.  

We consider the general case in which different initial conditions
are assigned to the different scalar VEV's $\langle Q_{i}\rangle =
\langle \overline{Q}_{i}^{\dagger}\rangle \equiv q_i$, and the 
system is described by $N_{f}$ coupled differential equations. 
Taking for illustration the case $N_{f}=2$, the equations to be 
solved are (see \cite{noi} for details)
\begin{eqnarray}
\ddot{q}_{1} + 3H\dot{q}_{1}-\frac{d\cdot q_{1}\ \Lambda ^{2a
}}{\left( q_{1}q_{2}\right) ^{2d N_{c}}} \left[
2+N_{c}\frac{q_{2}^{2}}{q_{1}^{2}}\right] &=& 0 \nonumber \; , \\
\ddot{q}_{2} + 3H\dot{q}_{2}- \frac{d\cdot q_{2}\ \Lambda ^{2a
}}{\left( q_{1}q_{2}\right) ^{2d N_{c}}} \left[
2+N_{c}\frac{q_{1}^{2}}{q_{2}^{2}}\right] &=& 0  \label{eom}
\end{eqnarray}
with $d = 1/(N_c -N_f)$ and $a= (3N_c-N_f)/(N_c-N_f)$.

In analogy with the one-scalar case, we look for power-law solutions 
of the form
\begin{equation}
q_{tr,i}=C_{i}\cdot t^{\, p_{i}}\ , \ \ i=1,\cdots ,\ N_{f}\ . 
\label{scaling}
\end{equation}
It is straightforward to verify that for fixed $N_f$ (and when 
$\rho _{Q}\ll \rho _{B}$), a solution exists with $p_i \equiv p =
p(N_c)$ and $C_i \equiv C =C(N_c,\Lambda)$ and is the same for all 
the $N_f$ flavors \cite{noi}.
The equation of state of the tracker is given by 
\begin{equation}
w_{Q}=\frac{1+r}{2}w_{B}-\frac{1-r}{2}\ , \label{eosfree}
\end{equation}
where we have defined $r \equiv N_{f}/N_c$.

Following the same methods employed in ref. \cite{scalcosmo} one can show
that this solution is the unique stable attractor in the space of
solutions of eqs. (\ref{eom}). Then, even if the $q_{i}$'s start with
different initial conditions, there is a region in field configuration
space such that the system evolves towards the equal fields solutions
(\ref{scaling}), and the late-time behavior is indistinguishable from
the case considered in ref. \cite{bin}, where equal initial conditions
for the $N_f$ flavors were chosen. 
In spite of this, the two-field dynamics introduces some new interesting 
features. For example, we have found that for any given initial energy
density such that -- for $q^{in}_1/q^{in}_2 =1$ -- the
tracker is joined before today, there exists always a limiting value for the
fields' difference above which the attractor is not reached in time.
A more detailed discussion and numerical results about the two-field 
dynamics can be found in \cite{noi}.

The scale $\Lambda $ can be fixed requiring that the scalar fields are
starting to dominate the energy density of the universe today and that
both have already reached the tracking behavior.  The two conditions
are realized if
\begin{equation}
v(q_{0})\simeq \rho _{crit}^{0}\ ,\ \ \ v^{\prime \prime
}(q_{0})\simeq H_{0}^{2}\ , \label{conditions}
\end{equation}
where $\rho _{crit}^{0}=3M_P^{2}H_{0}^{2}/8\pi $ and $q_{0}$ is 
the present scalar fields VEV. 
Eqs. (\ref{conditions}) imply
\[
\frac{\Lambda }{M_P} \simeq  \left[ 
\frac{3(1+r)(3+r)}{4\pi (1-r)^{2} rN_c} \right] ^{\frac{1+r}{2(3-r)}}
\!\!\left( \frac{1}{2rN_{c}}\frac{\rho _{crit}^{0}}{M_P^{4}}\right)
^{\frac{1-r}{2(3-r)}}
\]
\begin{equation}
\frac{q_{0}^{2}}{M_P^{2}} \simeq  \frac{3}{4\pi
}\frac{(1+r)(3+r)}{(1-r)^{2}} \frac{1}{rN_c} \; . \label{today}
\end{equation} 

\subsection{Interaction with the visible sector}

The superfields $Q_{i}$ and $\overline{Q}_{i}$ have been taken as
singlets under the SM gauge group. Therefore, they may interact with
the visible sector only gravitationally, {\it i.e.  }via
non-renormalizable operators suppressed by inverse powers of the
Planck mass, of the form
\begin{equation}
\int d^{4}\theta \ K^{j}(\phi _{j}^{\dagger },\phi _{j}) \cdot
\beta ^{ji}\left[ \frac{Q_{i}^{\dagger }Q_{i}}{M_P^{2}}\right] 
\; , \label{coupling}
\end{equation}
where $\phi _{j}$ represents a generic standard model superfield. From
(\ref{today}) we know that today the VEV's $q_{i}$ are typically
$O(M_P)$, so there is no reason to limit ourselves to the
contributions of lowest order in $|Q|^{2}/M_P^{2}$. Rather, we have
to consider the full (unknown) functions $\beta$'s
and the analogous $\overline{\beta }$'s for the $\overline{Q}_{i}$'s. 
Moreover, the requirement that the scalar fields are on the tracking
solution today, eqs. (\ref{conditions}), implies that their mass is of
order $\sim H_{0}\sim 10^{-33}$ eV. 

The exchange of very light fields gives rise to long-range forces
which are constrained by tests on the equivalence principle, whereas
the time dependence of the VEV's induces a time variation of the SM
coupling constants \cite{int}.  These kind of considerations set
stringent bounds on the first derivatives of the $\beta ^{ji}$'s {\it today,}
\[
\alpha ^{ji}\equiv \left.\frac{d\log \beta ^{ji}\left[ x_i^2 \right]
}{d x_i}\right|_{x_i=x_i^0} ,
\]
where $x_i \equiv q_i/M_P$.  To give an example, the best bound on the
time variation of the fine structure constant comes from the Oklo
natural reactor. It implies that $\left| \dot{\alpha}/\alpha \right|
<10^{-15}\ {\rm yr}^{-1}$ \cite{dam2}, leading to the following
constraint on the coupling with the kinetic terms of the
electromagnetic vector superfield $V$,
\begin{equation}
\alpha ^{Vi}\ \lta\ 10^{-6}\ \frac{H_{0}}{
\left\langle \dot{q}_{i}\right\rangle }\,M_P \,, \label{decoupling}
\end{equation}
where $\left\langle \dot{q}_{i}\right\rangle $ is the average rate of
change of $q_{i}$ in the past $2\times 10^{9}{\rm yr}$. 
Therefore, in order to be phenomenologically
viable, any quintessence model should postulate that all the unknown
couplings $\beta ^{ji}$'s and $\overline{\beta }^{ji}$'s have a
common minimum close to the actual value of the $q_{i}$'s.

The simplest way to realize this condition would be via the {\it least
coupling principle } introduced by Damour and Polyakov for the
massless superstring dilaton in ref. \cite{dam3}, where a universal
coupling between the dilaton and the SM fields was postulated. In the
present context, we will invoke a similar principle, by postulating
that $\overline{\beta} ^{ji}=\beta ^{ji}=\beta $ for any SM 
field $\phi _{j}$ and any flavor $i$.

The decoupling from the visible sector implied by bounds like (\ref
{decoupling}) does not necessarily mean that the interactions between
the quintessence sector and the visible one have always been
phenomenologically irrelevant. Indeed, during radiation domination the
VEVs $q_{i}$ were typically $\ll M_P$ and then very far from the
postulated minimum of the $\beta $'s. For such values of the
$q_{i}$'s the $\beta $'s can be approximated as
\begin{equation}
\beta \left[ \frac{Q^{\dagger }Q}{M_P^{2}}\right] =\beta _{0}+\beta
_{1}\frac{Q^{\dagger }Q}{M_P^{2}}\ +\ldots \label{betarad}
\end{equation}
where the constants $\beta _{0}$ and $\beta _{1}$ are not directly
constrained by (\ref{decoupling}).  The coupling between the
last expression and the SM\ kinetic terms, as in (\ref{coupling}),
induces a SUSY breaking mass term for the scalars of the form
\begin{equation}
\Delta L\sim H^{2}\,\beta_1 \sum_{i}\ (\left| Q_{i}\right| ^{2}+\left|
\overline{Q}_{i}\right| ^{2}) \label{masses}
\end{equation}
as discussed in \cite{drt}.

If present, this term would have a very interesting impact on the
cosmological evolution of the fields. 
From a phenomenological point of view, the most relevant effect of the
presence of mass terms like (\ref{masses}) during radiation domination
is the rise of the scalar potential at large
fields values, that induces a (time-dependent) minimum. 
This results in a significative enlargment of the already large region
of initial condition phase space leading to late-time tracking behavior.
Numerical confirmation of this qualitative discussion can be found in 
\cite{noi}.

\section{INFLATION}

The idea of studying inflaton potentials $V(\phi)$ which go 
to zero at infinity is not new \cite{quinf,pv,no} and has most 
recently referred to as the ``quintessential inflation'' \cite{pv} or 
``non oscillatory'' \cite{no} scheme. 
All these models have the appealing feature of providing a natural 
candidate for quintessence in the tail of the inflaton potential.

The main emphasis of these previous works was on the mechanism of
rehating which, due to the unsual shape of the potential, could not
be achieved in the standard oscillatory way.
Gravitational particle production was most often invoked
\cite{quinf,pv}, but Felder, Kofman and Linde have recently shown 
\cite{no} that the so-called ``instant preheating'' mechanism
is also a workable option.
We will instead focus on the issue of the compatibility of the
constraints coming from inflation and quintessence \cite{noi2}.

Regarding {\it inflation}, there are four main points to be taken 
into account:

\noindent
1. The equation of state of the inflaton $\phi$ must satisfy  
$w_{\phi} < -1/3$, for the universe to accelerate. 

\noindent
2. A sufficient number of e-foldings should take place,
in order to solve the flatness and horizon problems. 

\noindent
3. The amplitude of scalar perturbations in the
cosmic microwave background measured by COBE constrains the 
normalization of the inflaton potential.

\noindent
4. We must ensure that at the end of inflation 
sufficient reheating takes place. 

For what concerns {\it quintessence}, instead, the following 
requirements should be fulfilled:

\noindent
1. In order for the scalar field modeling of the cosmological constant
to be sufficiently general, we require that the late-time shape of the
potential is of the form
$V(\phi) \sim M^{4+\alpha} \phi^{-\alpha}$ with $\alpha > 0$.

\noindent
2. Secondly, we want the field $\phi$ to be already on track today
and its present energy density to correspond to what observations report, 
i.e. $\Omega_{\phi} \simeq 2/3$, as discussed in Section 2.

While it is straightforward to find potentials with the required early 
and late-time behavior, the subtle issue resides in successfully matching 
the exit conditions for the scalar field after inflation with the range 
of initial conditions allowed for the trackers. 

An very promising possibility seems to be that of considering
first-order inflation. In this case, if the potential does not have an
absolute minimum but instead goes to zero at infinity, the exit
conditions of the inflaton $\phi$ from the tunneling would set to a very high
precision the initial conditions for the subsequent quintessential
evolution of the same field $\phi$.
In ref. \cite{noi2}, we study the scalar field dynamics in a potential
of the form
\begin{equation}
V \left( \phi \right) =
\frac{\Lambda^{\alpha+6}}{\phi^{\alpha} \left[ \left( \phi -
v \right)^2 + \beta^2 \right]} \;\; , \;\; \mbox{with} 
\;\;\frac{\beta}{v}\ll 1 \; ,
\end{equation}
where $\Lambda$, $\beta$ and $v$ are constants of mass dimension one.
This potential has a barrier  in $\phi \simeq v $, 
while for $\phi \gg v$ it behaves like 
$V \left( \phi \right) \sim \Lambda^{\alpha+6} \phi^{-\alpha -2}$ 
as required by the tracker condition.
Reheating can be easily achieved via bubble collisions after
nucleation, as it is usually done in first order inflation.

\vskip.2in
I would like to thank Antonio Masiero, Massimo Pietroni and Marco
Peloso with whom the results reported here were obtained. I am also
grateful to Antonio Riotto for a number of enlightening discussions.


\begin{thebibliography}{9}

%
%
\bibitem{data}
 N.A.~Bahcall, J.P.~Ostriker, S.~Perlmutter and P.J.~Steinhardt, 
 astro-ph/9906463.

%
\bibitem{quint} 
 J.A.~Frieman and I.~Waga, Phys.  Rev. {\bf D57}, 4642 (1998); 
 R.R.~Caldwell, R.~Dave, and P.J.~Steinhardt, Phys. Rev. Lett. {\bf 80}, 1582 
 (1998); 
 M.S.~Turner and M.~White, Phys. Rev. {\bf D56}, 4439 (1997).

%
\bibitem{bin} 
 P.~Bin\'{e}truy, hep-ph/9810553. 
\bibitem{noi}
 A.~Masiero, M.~Pietroni and F.~Rosati, hep-ph/9905346. 
\bibitem{mod2} 
 J.A.~Frieman, C.T.~Hill, A.~Stebbins, and I.~Waga, Phys. Rev. Lett. {\bf 75}, 
 2077 (1995); 
 K.~Choi,  hep-ph/9902292;  
 J.E.~Kim,  hep-ph/9811509;
 P. Brax and J. Martin, astro-ph/9905040;
 M.C.~Bento and O.~Bertolami, gr-qc/9905075.

%
\bibitem{det} 
 L.~Wang, R.R.~Caldwell, J.P.~Ostriker, and P.J.~Steinhardt, 
 astro-ph/9901388; 
 G.~Huey, L.~Wang, R.~Dave, R.R.~Caldwell, and P.J.~Steinhardt, 
 Phys. Rev {\bf D59}, 063005 (1999); 
 S.~Perlmutter, M.S.~Turner, and M.~White, astro-ph/9901052; 
 T. Chiba, N. Sugiyama and T. Nakamura, MNRAS vol. {\bf 301} 72 (1998);
 D.~Huterer and  M.S.~Turner, astro-ph/9808133;
 T. Nakamura, T. Chiba,  astro-ph/9810447;
 T. Chiba, T. Nakamura, Prog.Theor.Phys. {\bf 100} 1077 (1998) . 

%
\bibitem{scalcosmo} 
 P.J.E.~Peebles and  B.~Ratra, Astrophys. Jour. {\bf 325}, L17 (1988); 
 B.~Ratra and P.J.E.~Peebles, Phys. Rev. {\bf D37}, 3406 (1988);
 A.R.~Liddle and R.J.~Scherrer, Phys. Rev. {\bf D59}, 023509 (1999).

%
\bibitem{scale} 
 C.~Wetterich, Nucl. Phys. {\bf B302}, 668 (1988);
 E.J.~Copeland, A.R.~Liddle, and D.~Wands, Phys.Rev. {\bf D57}, 4686 (1998);
 P.G.~Ferreira and M.~Joyce, Phys. Rev.  Lett. {\bf 79}, 4740 (1997);  
 P.G.~Ferreira and M.~Joyce, Phys. Rev. {\bf D58}, 023503 (1998). 
 
%
\bibitem{track} 
 I.~Zlatev, L.~Wang, and P.J.~Steinhardt, Phys. Rev. Lett. {\bf 82}, 
 896 (1999);
 P.J.~Steinhardt, L.~Wang, and I.~Zlatev, Phys. Rev. {\bf D59}, 123504 (1999). 


\bibitem{int} 
 S.M.~Carroll, Phys. Rev.  Lett.  {\bf 81}, 3067 (1998);
 T.~Damour, gr-qc/9606079. 

%
\bibitem{SQCD} 
 T.R.~Taylor, G.~Veneziano, and S.~Yankielowicz, Nucl. Phys. B {\bf 218}, 493 
 (1983); 
 I.~Affleck, M.~Dine, and N.~Seiberg, Phys. Rev. Lett. {\bf 51}, 1026 (1983), 
 Nucl.  Phys. B {\bf 241}, 493 (1984). 
 See also: M.E.~Peskin,  hep-th/9702094. 

%
\bibitem{dam2} 
 T.~Damour and F.~Dyson, Nucl.\ Phys. {\bf B480}, 37 (1996). 
\bibitem{dam3} 
 T.~Damour and A.M.~Polyakov, Nucl.\ Phys. {\bf B423}, 532 (1994). 
\bibitem{drt} 
 M.~Dine, L.~Randall, and S.~Thomas, Phys. Rev. Lett. {\bf 75}, 398 (1995). 

%
\bibitem{quinf}
 L.H.~Ford, Phys. Rev. {\bf D35}, 2955(1987); 
 B.~Spokoiny, Phys. Lett. {\bf B315}, 40 (1993); 
 M.~Joyce and T.~Prokopec, Phys. Rev. {\bf D57}, 6022 (1998).
\bibitem{pv}
 P.J.E.~Peebles and A.~Vilenkin, Phys.Rev. {\bf D59}, 063505 (1999).
\bibitem{no} 
 G.~Felder, L.~Kofman and A.~Linde,  hep-ph/9903350.

\bibitem{noi2}
 M.~Peloso and F.~Rosati, hep-ph/9908271.

\end{thebibliography}
\end{document}